\documentclass{article}
\newcommand{\be}{\begin{equation}}
\newcommand{\ee}{\end{equation}}
\def\bea{\begin{eqnarray}}
\def\eea{\end{eqnarray}}
\def\bean{\begin{eqnarray*}}
\def\eean{\end{eqnarray*}}

\def\bea{\begin{eqnarray}}
\def\eea{\end{eqnarray}}
\def\sla{\raise.15ex\hbox{$/$}\kern-.57em}

\usepackage{amsmath}
\usepackage{latexsym}
\newcommand{\deltab}{\boldsymbol\delta}
\begin{document}

\begin{center}
{\bf \Large STILL IN LIGHT-CONE SUPERSPACE}
\end{center}
\vskip .2cm

\begin{center}{P. Ramond\footnote{In collaboration with D. Belyaev, L. Brink, and S-S. Kim}}
{Physics Department, \\Institute for Fundamental Theory,\\University of Florida,\\
Gainesville, FL 32611,  USA\\
E-mail: ramond@phys.ufl.edu
}\end{center}

\begin{abstract}
\noindent The recently formulated Bagger-Lambert-Gustavsson (BLG)  
theory in three dimensions is described in terms of a constrained chiral superfield in light-cone superspace. We discuss the use of Superconformal symmetry to determine the form of its interactions, in complete analogy with $N=4$ SuperYang-Mills in four dimensions.  
\end{abstract}

%\keywords{ Superspace; Light-cone; Superconformal Theories}

%\bodymatter

\section{Introduction}
Maximally supersymmetric theories live in two different superspaces. The first with eight complex Grassmann variables is used to describe $N=1$ supergravity in $d=11$, $N=8$ supergravity in $d=4$,  $N=16$ Supergravity in $d=3$, and so-on.  With a dimensionful coupling, these theories are not superconformal. Instead they contain non-compact and non-linear symmetries, $E_{7(7)}$ and $E_{8(8)}$ in $d=4$ and $3$, respectively. 

The second superspace with only four complex Grassmann variables is equally rich. It houses theories with Superconformal symmetry in $d=6,5,4$ and $3$ dimensions. The latter theory has been recently formulated covariantly \cite{BAG,GUS}{}, and on the light-cone \cite{BENGT}{}.  

It has already been shown\cite{ANANTH}  how the fully interacting $N=4$ SuperYang-Mills theory\cite{N=4}  in $d=4$ can be determined by requiring $PSU(\,2,2\,|\,4\,)$ Superconformal symmetry on a constrained chiral superfield in light-cone superspace with four complex Grassmann variables. 

The following is a progress report on using the same technique, now applied to $OSp(\,2,2\,|\,8\,)$ Superconformal symmetry on the same chiral superfield. On the light-cone, supersymmetries split into kinematical and dynamical supersymmetries. Kinematical supersymmetries are linearly realized on the chiral superfield. The dynamical ones also contain a linear term (free theory), but also terms non-linear in the (super)fields, which, in superconformal theories,  suffice to {\it completely} determine the theory. Our technique has been to use algebraic consistency to find its expression. 

Consistency with the kinematical constraints yields two possible expressions for the dynamical supersymmetries, each determined in terms of four integers, and with an unknown four-index tensor $f^{abcd}$, where the indices label the superfields. The values of these  integers are determined by requiring that the light-cone Hamiltonian and boosts commute with one another.  At Shifmania (this proceeding), we reported an unexpected solution, with fractional light-cone derivatives acting on the chiral superfield, without assuming any symmetry among the indices of $f^{abcd}$.  Since then, we have found\cite {US} that by requiring antisymmetry in three of its indices, $b\leftrightarrow c$, $b\leftrightarrow d$ and $c\leftrightarrow d$, the BLG solution emerges from these algebraic constraints, apparently uniquely. 

%%%%%%%%%%%%%%%%%
\section{Superconformal Theories}
In 1978, W. Nahm\cite{NAHM}  catalogued all relativistic field theories which extended the Poincar\'e symmetry to Superconformal symmetry. We only list those in  spacetime dimensions $d=6,5,4,$ and $3$: 

\bean
d=6&& OSp(\,2n\,|\,6,2\,)~\supset~SO(6,2)\times Sp(2n)^{}_R\ ,\\
d=5&& F[4]~\supset~SO(5,2)\times SU(2)^{}_R\ ,\\
d=4&& SSU(\,2,2\,|\,n\,)~\supset~SO(4,2)\times SU(n)^{}_R\times U(1)^{}_R\ ,\\
d=4&& PSU(\,2,2\,|\,4\,)~\supset~SO(4,2)\times SU(4)^{}_R\ ,\\
d=3&& OSp(\,2,2\,|\,n\,)~\supset~SO(3,2)\times SO(n)^{}_R\ ,\\
\eean
using Kac's notation for the superalgebras. The conformal group in $d$ spacetime dimensions is $SO(d,2)$; for $d=4$, it is a non-compact form of $SU(2,2)$, and for $d=3$ it is isomorphic to $Sp(2,2)$. These theories have large global $R$-symmetries.  The theories with special number of $R$-symmetries, $n=2$ in $d=6$, $n=4$ in $d=4$, and $n=8$ in $d=3$, can be described in terms of constrained chiral superfields in light-cone superspace. Since then, it has been realized that many superconformal theories are seminal, not only in quantum field theory but also in Superstrings and M-theory\cite{MANY}{}.

%%%%%%%%%%%%%%%
\section{$N=4$ Light-Cone Superspace}
We introduce the usual light-cone variables

\be
x^\pm_{}=\frac{1}{\sqrt{2}}(x^0_{}\pm x^3_{})\ ,\quad \partial^\pm_{}=\frac{1}{\sqrt{2}}(\partial^0_{}\pm \partial^3_{})\ ,
\ee
and denote the transverse variables by $x_1...x_{d-2}$. The relevant  superspace contains four complex Grassmann variables, $\theta^m$ and $\bar\theta_m$, in terms of which we define the chiral derivatives

\be
d^{m}_{}~=~-\frac{\partial}{\partial\bar\theta_m}\,-\,\frac{i}{\sqrt2}\,\theta^m\,\partial^+\ ;\qquad \bar d_{n}~=~ \frac{\partial}{\partial\theta_n}\,+\,\frac{i}{\sqrt2}\,\bar\theta_n\,\partial^+\ ;
\nonumber\ee
they  satisfy 

\be
\{\,d^m_{}\,,\,\bar d_{n}\,\}~=-i\,\sqrt2\,\delta^m_n\,\partial^+\ .
\ee
The chiral superfields

\bean
\varphi^{\,a}_{}\,(y)&=&\frac{1}{ \partial^+}\,A^{\,a}_{}\,(y)\,+\,\frac{i}{\sqrt 2}\,{\theta_{}^m}\,{\theta_{}^n}\,{\overline C^{\,a}_{mn}}\,(y)\,+\,\frac{1}{12}\,{\theta_{}^m}\,{\theta_{}^n}\,{\theta_{}^p}\,{\theta_{}^q}\,{\epsilon_{mnpq}}\,{\partial^+}\,{\bar A}^{\,a}_{}\,(y)\cr
& &~~~ +~\frac{i}{\partial^+}\,\theta^m_{}\,\bar\chi^{\,a}_m(y)+\frac{\sqrt 2}{6}\theta^m_{}\,\theta^n_{}\,\theta^p_{}\,\epsilon^{}_{mnpq}\,\chi^{q\,a}_{}(y)
\eean
where $a$ is a taxonomic  index, are chiral by construction,  

\be
d^m_{}\,\varphi^a\,(y)~=~0\ ,
\ee
where the component fields depend on the chiral coordinates  

\be
y~=~(\,x^{}_1,...,x^{}_{d-2}\,,\,x^--i\frac{\theta^m\bar\theta_m}{\sqrt{2}}\,).
\ee
The parameter $x^+$ is set to zero without loss of generality. The chiral superfields  obey the ``inside-out" constraint  

\be
\overline d^{}_m\,\overline d^{}_n\,\varphi^a~=~\frac{1}{2}\,\epsilon^{}_{mnpq}\,d^p_{}\,d^q_{}\,\overline\varphi^a\ .
\ee
In  $d=4$ SuperYang-Mills, this important constraint allowed us\cite{ANANTH} to write its light-cone interacting Hamiltonian as a {\em positive definite} quadratic form.  

The component fields of each chiral superfield represent sixteen physical degrees of freedom, eight bosons and eight fermions. They are organized in terms of an $SO(8)$ $R$-symmetry, with the bosons transforming as a vector, the fermions as a spinor. 

Introduce the operators

\be
q^{m}_{}~=~-\frac{\partial}{\partial\bar\theta_m}\,+\,\frac{i}{\sqrt2}\,\theta^m\,\partial^+\ ;\quad \bar q_{n}~=~ \frac{\partial}{\partial\theta^n}\,-\,\frac{i}{\sqrt2}\,\bar\theta_n\,\partial^+\ ,
\ee
which satisfy  

\be
\{\,q^m_{}\,,\,\bar q_{n}\,\}~=~i\,\sqrt2\,\delta^m_n\,\partial^+\ ,
\ee
and do not alter chirality, since they anticommute with the chiral derivatives

\be
\{\,q^m_{}\,,\,\bar d^{}_n\,\}~=~\{\,q^m_{}\,,\, d^{n}_{}\,\}~=~0\ .
\ee 
The $SO(8)$ transformations are written in terms of those of its $SO(6)\times U(1)\sim SU(4)\times U(1)$ subgroup, with parameters  $\omega^{m}_{~~n}$,  and 
$\omega$: 

\be\label{SO8-1}
\delta^{}_{SO(6)}\,\varphi^a_{}~=~\omega^m_{~~n}\,\frac{i}{\sqrt{2}}\left( q^n_{}\,\bar q_m^{}-\frac{1}{4}\delta^n_{\,m}\,q^l_{}\,\bar q_l^{}\,\right)\frac{1}{\partial^+_{}}\,\varphi^a_{}\ ; 
\ee
\be\label{SO8-2}
\delta^{}_{U(1)}\,\varphi^a_{}~=~\omega\,\frac{i}{4\sqrt{2}}\left( q^m_{}\,\bar q_m^{}-\bar q_m^{}\, q^m_{}\,\right)\frac{1}{\partial^+_{}}\,\varphi^a_{}\ ; 
\ee
and the coset parameters $\omega^{mn}$, and $\overline\omega_{mn}$, 

\be\label{SO8-3}
\delta^{}_{\overline{coset}}\,\varphi^a_{}~=~\omega^{mn}_{}\,\frac{i}{\sqrt{2}}\, \bar q^{}_m\,\bar q_n^{}\,\frac{1}{\partial^+_{}}\,\varphi^a_{}\ ; \quad
\delta^{}_{{coset}}\,\varphi^a_{}~=~{\omega}^{}_{mn}\,\frac{i}{\sqrt{2}}\,  q^{m}_{}\, q_{}^n\,\frac{1}{\partial^+_{}}\,\varphi^a_{}\ , 
\ee

This chiral superfield can be used to define theories in different dimensions, with the only modifications of increase the number of transverse coordinates of its component fields:  
%In ten spacetime dimensions, $SO(8)$ is the transverse little group, and as the number of space dimensions decreases, so does the size of the little group, and the remnant is interpreted as $R$-symmetries. 

\begin{itemize}

\item $d=10$

The superfield describes $N=1$ in $d=10$ dimensions. This theory is not superconformal, as it is the zero slope limit of an open superstring. The $SO(8)$ transformations are interpreted as the ``spin" part of the transverse little group, the orbital part being supplied by the appropriate number of transverse coordinates. There are no modifications to  the chiral superfield, except for the dependence of its components on the six extra transverse coordinates.

\vskip .2cm
\item $d=6$  

The superconformal group is $OSp(\,4\, | \,6,2\,)$. The transverse light-cone little group is $SO(4)\sim SU(2)\times SU(2)$. The first $SU(2)$ has only an orbital part, whereas the spin part of the second $SU(2)$ is to be found in the decomposition

\be
SO(8)~\supset~SU(2)\times Sp(4)_R\ ,
\ee
where the physical fields decompose as

\be
{\bf 8}_b~=~(\,{\bf 3}\,,\,{\bf 1}\,)+(\,{\bf 1}\,,\,{\bf 5}\,)\ ,\qquad {\bf 8}_f~=~(\,{\bf 2}\,,\,{\bf 4}\,)\ .\ee
The bosons split into an $R$-quintet of scalar fields and an $R$-singlet tensor, a second rank antisymmetric tensor with self-dual three-form field strength.  

\vskip .2cm
\item $d=5$

The superconformal symmetry group is $F[4]$. The transverse little group is $SO(3)\sim SU(2)$, and its spin part is to be found in the decomposition 

\be
SO(8)~\supset~SU(2)\times SU(2)_R\ .
\ee
The $R$-symmetry reduces to $SU(2)$. This decomposition is similar to that in $d=6$,  with the anomalous embedding of $SU(2)$ in $Sp(4)$ with

\be
Sp(4)~\supset~ SU(2)\ ,\qquad {\bf 5}~=~{\bf 5}\ ,\quad{\bf 4}~=~\bf{4}\ ,\ee
so that the scalar bosons split into one $R$-singlet vector, and five scalars with $R$-spin $2$ and the fermions $R$-spin $3/2$.

\vskip .2cm
\item $d=4$
 
The little group is now just $SO(2)$ whose spin part is found in 

 \be
 SO(8)~\supset~SO(2)\times SO(6)_R\sim U(1)\times SU(4)_R\ .
 \ee
This leads to the well-known $N=4$ SuperYang-Mills theory, symmetric under $PSU(\,2,2\,|\,4\,)$, with one vector and six scalars.  

\vskip .2cm
\item $d=3$

There is no light-cone little group, and the $R$-symmetry is the full $SO(8)$. The chiral superfield describes the degrees of freedom in the 
Nahm theory with $n=8$, and symmetry $OSp(\,2,2\,|\,8\,)$. The bosons (fermions) form an $R$-symmetry vector (spinor) octet.  

The light-cone formulation of this theory will occupy the rest of this paper, using 
algebraic techniques previously developed for the $N=4$ theory in four dimensions. 

\end{itemize}

%%%%%%%%%%%%%%%%%%%%%%%
\section{$OSp(2,2\,|\,8)$ Generators}
We begin with 

\[
OSp(2,2\,|\,8)\supset Sp(2,2)\times SO(8)\ ,\]
where the first factor group is the conformal group in three dimensions, and the second factor group is the $R$-symmetry. 

In light-cone coordinates, the space-time generators are either  kinematical or dynamical. The kinematical generators operate within the initial surface ($x^+=0$), while the dynamical generators, called hamiltonians by Dirac, act transversely to the initial surface. The kinematical operators are the same in free and interacting theories, and are linear in the (super)fields. The dynamical operators also contain a part linear in the (super)fields for the free theory, but in the interacting theory, they develop non-linear dependence on the (super)fields. 

The ten generators of the conformal group in three dimensions are given by  

$$\rm Conformal ~Group~~~\begin{cases}
~~{\rm  Lorentz~ Group: }\quad J^{+-}\ , J^+ \ ;\quad \mathcal{J}^-\\
~~{\rm  Translations: }\quad P\ , P^+\ ;\quad \mathcal{P}^-\\
~~{\rm  Dilatation: }\quad D \\
~~{\rm  Conformal: }\quad  K\ ,  K^+\ ;\quad \mathcal{K}^-
\end{cases}\ ,$$
with the dynamical generators written in capital calligraphic letters. Note that $J^+$ and $K^+$ can be viewed as kinematical as long as we set the parameter $x^+$ to zero. 

The supersymmetry and superconformal generators

$$\rm Supers~~~\begin{cases}
~~{\rm  Supersymmetry: }\quad q\ , \bar q\ ;\quad {\mathcal Q}\ , \overline{\mathcal Q}\\
~~{\rm  Superconformal: }\quad s\ , \bar s\ ;\quad {\mathcal S}\ , \overline{\mathcal S}
\end{cases}\ ,$$
also split into kinematical and dynamical operators.  All $R$-symmetry generators are kinematical, and given by Eqs.(\ref{SO8-1}-\ref{SO8-3}).

%%%%%%%%%%%%%%%%%%%%%%%%%%
\subsection{Kinematical Transformations} 
They  are expressed in terms of 

\be
{\cal N}~=~\theta^m_{}\frac{\partial}{\partial\theta^m_{}}\,+\,\bar\theta^{}_m\frac{\partial}{\partial\bar\theta^{}_m}\ ;\qquad
{\cal A}~\equiv~x^-\,\partial^+-\frac{x}{2}\,\partial\,-\frac{1}{2}{\cal N}+\frac{1}{2}\ .
\ee
The kinematical Poincar\' e  transformations are

\be
\delta^{}_{P^{+}}\,\varphi^a_{}~=-i\,\partial^+_{}\,\varphi^a_{}\ ;\qquad \delta^{}_{P}\,\varphi^a_{}~=-i\,\partial\,\varphi^a_{}\ ;\ee

\be
\delta^{}_{J^{+}}\,\varphi^a_{}~=~ix\,\partial^+_{}\,\varphi^a_{}\ ;\qquad 
\delta^{}_{J^{+-}}\,\varphi^a_{}~=~i(\,x^-\,\partial^+_{} -\frac{1}{2}{\cal N}\,+\,1\,)\,\varphi^a_{}\ ; \ee
 
\be
\delta^{}_{P^+}\,\varphi^a_{}~=~-i\partial^+_{}\,\varphi^a_{}\ ; \qquad
\delta^{}_{P}\,\varphi^a_{}~=~-i\partial\,\varphi^a_{}\ , 
\ee
followed by the kinematical conformal symmetries 

\bea
\delta^{}_{D}\,\varphi^a_{}&=&i\,(\,x^-\partial^+\,-\,x\,\partial\,-\frac{1}{2}{\cal N}\,+\frac{1}{2}\,)\,\varphi^a_{}\ ;\\
\delta^{}_{K}\,\varphi^a_{}&=&2i\,x\,\mathcal A\,\varphi^a_{} \ ;\qquad
\delta^{}_{K^+}\,\varphi^a_{}~=i\,x^2\partial^+_{}\,\varphi^a_{}\ . 
\eea
Similarly, the kinematical (spectrum generating) supersymmetries, with parameters $\varepsilon^m$ and $\bar\varepsilon_m$, are  

\be
\delta^{\,kin}_{\varepsilon\bar q}\,\varphi^a_{}~=~\varepsilon^m_{}\bar q_m^{}\,\varphi^a_{}\ ; \qquad \delta^{\,kin}_{\bar\varepsilon q}\,\varphi^a_{}~=~\bar\varepsilon^{}_m q^m_{}\,\varphi^a_{}\ ,
\ee
and finally kinematical  superconformal transformations with parameters $\alpha^m$ and $\bar\alpha_m$

\be
\delta^{}_{\alpha\bar s}\,\varphi^a_{}~=-ix \alpha_{}^m\,\bar q_m^{}\,\varphi^a_{}\ ;\qquad
\delta^{}_{\bar\alpha s}\,\varphi^a_{}~=~ix\, \bar\alpha^{}_m\,q^m_{}\,\varphi^a_{}\ .
\ee
%%%%%%%%%%%%
\subsection{Free Dynamical Transformations}
In superconformal theories, {\bf all} dynamical generators are determined by the algebra from the dynamical supersymmetry transformations, because the algebra is simple. To see how this works, we start from the {\em free} dynamical supersymmetry transformations (written in bold), which are given by

\be
\deltab^{free}_{\varepsilon\overline {\mathcal Q}}\,\varphi^a~=~\frac{1}{\sqrt{2}}\varepsilon_{}^m\bar q^{}_m\,\frac{\partial}{\partial^+_{}}\,\varphi^a_{}\ ,\qquad 
\deltab^{free}_{ \bar\varepsilon{\mathcal Q}}\,\varphi^a~=~\frac{1}{\sqrt{2}}\bar\varepsilon^{}_m q^{m}_{}\,\frac{\partial}{\partial^+_{}}\,\varphi^a_{}\ .\ee
We then use the commutators 

\be
[\,\deltab^{}_{\varepsilon\overline {\mathcal Q}}\,,\,\deltab^{}_{\bar\varepsilon{\mathcal Q}}\,]\,\varphi^a_{}~=~\sqrt{2}\,\bar\varepsilon^{}_m\varepsilon_{}^m\,\deltab^{}_{\mathcal{P}^-}\,\varphi^a_{}~~~\rightarrow~~~\deltab^{}_{\mathcal{P}^-}\,\varphi^a_{}\ ,\ee
\be
[\,\delta^{}_{K}\,,\,\deltab^{}_{\mathcal{P}^-}\,]\,\varphi^a_{}~=~{2i}\,\deltab^{}_{\mathcal{J}^-}\,\varphi^a_{}~~~~~~~~~~~~~\rightarrow~~~\deltab^{}_{\mathcal{J}^-}\,\varphi^a_{}\ ,\ee
\be
[\,\delta^{}_{K}\,,\,\deltab^{}_{\mathcal{J}^-}\,]\,\varphi^a_{} ~~~=~-i\,\deltab^{}_{\mathcal{K}^-}\,\varphi^a_{} ~~~~~~~~~\rightarrow~~~\deltab^{}_{\mathcal{K}^-}\,\varphi^a_{}\ ,\ee
\be
[\,\delta^{}_{K}\,,\,\deltab^{}_{\varepsilon\overline{\mathcal Q}}\,]\,\varphi^a_{}~=~\sqrt{2}\,\deltab^{}_{\varepsilon\overline{\mathcal S}}\,\varphi^a_{}~~~~~~~~~~~~~\rightarrow~~~\deltab^{}_{\varepsilon\overline{\mathcal S}}\,\varphi^a_{}\ ,
\ee
to compute the remaining dynamical transformations. Evaluation of the commutators yields 

\bean
{\rm Time~Translation:}&&~~\deltab^{free}_{\mathcal{P}^{-}_{}}\,\varphi^{a}_{}
~=-i\,\frac{\partial^2_{}}{2\,\partial^+}\,\varphi^a_{}\ ,\\ 
{\rm Lorentz~ Boost:}&&~~\deltab^{free}_{\mathcal{J}^{-}}\,\varphi^{a}_{}
~=~-i\frac{\partial}{\partial^+}{\cal A}\,\varphi^a_{} \ ,
\\
{\rm Conformal~ Boost:}&&~~\deltab^{free}_{\mathcal{K}^{-}}\,\varphi^{a}_{}~=~2i\,\frac{1}{\partial^+}\,{\cal A}\,({\cal A}-\frac{1}{2})\,\varphi^a_{} \ ,
\\
{\rm SuperConformal:}&&~~\deltab^{free}_{\alpha\overline  {\mathcal S}}\,\varphi^a~=~i\,\alpha^m_{}\bar q^{}_m\,\frac{1}{\partial^+_{}}\,\mathcal{A}\,\varphi^a_{}\ ,\\ 
&&~~ \deltab^{free}_{ \bar \alpha{\mathcal S}}\,\varphi^a~=-i\,\bar\alpha^{}_m q^m_{}\,\frac{1}{\partial^+_{}}\,\mathcal{A}\,\varphi^a_{}\ .
\eean
These are valid in the free theory, and need to be altered in the interacting theory.

%%%%%%%%%%%%%%%%%%%%%%%%%
\subsection{Interacting Dynamical Supersymmetries}
Just as in the free case, it suffices to determine the form of the dynamical supersymmetry transformations. We write 

\be
\deltab^{}_{\varepsilon\overline {\mathcal Q}}\,\varphi^a~=~\deltab^{free}_{\varepsilon\overline {\mathcal Q}}\,\varphi^a+\deltab^{ int}_{\varepsilon\overline {\mathcal Q}}\,\varphi^a\ ,\qquad \deltab^{}_{\overline\varepsilon {\mathcal Q}}\,\varphi^a~=~\deltab^{free}_{\overline\varepsilon {\mathcal Q}}\,\varphi^a+\deltab^{ int}_{\overline\varepsilon {\mathcal Q}}\,\varphi^a\ .
\ee
The expressions $\deltab^{int}_{\varepsilon\overline {\mathcal Q}}\,\varphi^a$ and $\deltab^{ int}_{\overline\varepsilon {\mathcal Q}}\,\varphi^a$are  highly restricted, by the following ten constraints:

\begin{enumerate}

\item Chirality 

\be
d^m\,\deltab^{int}_{\varepsilon\overline {\mathcal Q}}\,\varphi^a~=~d^m\,\deltab^{int}_{\overline\varepsilon {\mathcal Q}}\,\varphi^a~=~0\ .\ee

\item Both $\deltab^{int}_{\varepsilon\overline {\mathcal Q}}\,\varphi^a$ and $\deltab^{int}_{\overline\varepsilon {\mathcal Q}}\,\varphi^a$ are cubic in the superfields. 

In three dimensions, canonical Bose fields have mass dimension of one-half, so that the chiral superfield has half-odd integer canonical dimension itself, assuming integer power of derivatives. Since we are looking for a conformal theory with no dimensionful parameters, $\deltab^{}_{\varepsilon\overline{\mathcal Q} }\,\varphi^a$ and $\deltab^{}_{\overline\varepsilon{\mathcal Q} }\,\varphi^a$ must then both be an odd power of superfields. 
Also, conformal invariance requires a Hamiltonian with a local sixth-order interaction in the superfields: the non-linear part of the dynamical supersymmetry transformation must be {\em cubic} in the superfields\cite{BASU}{}: the theory must have a tensor with four indices\footnote{In $d=4$, similar considerations suggested a tensor with three indices, $f^{abc}$, which turned out to be the structure functions of the gauge algebra.}.

\item Both are 
%$\deltab^{int}_{\varepsilon\overline {\mathcal Q}}\,\varphi^a$ is 
independent of $x^-$, using

\be
[\,\delta^{}_{P^+}\,,\,\deltab^{}_{\varepsilon\overline {\mathcal Q}}\,]\,\varphi^a~=~[\,\delta^{}_{P^+}\,,\,\deltab^{}_{\overline\varepsilon {\mathcal Q}}\,]\,\varphi^a~=~0
\ .\ee

\item $\deltab^{int}_{\varepsilon\overline {\mathcal Q}}\,\varphi^a$ is independent of $x$, since

\be
[\,\delta^{}_{P}\,,\,\deltab^{}_{\varepsilon\overline {\mathcal Q}}\,]\,\varphi^a~=~0\ .
\ee

\item Neither have
%$\deltab^{int}_{\varepsilon\overline {\mathcal Q}}\,\varphi^a$ has no 
transverse derivatives $\partial$: from

\[
[\,\delta^{}_{J^+}\,,\,\deltab^{}_{\varepsilon\overline {\mathcal Q}}\,]\,\varphi^a~=-\frac{i}{2}\,\deltab^{}_{\varepsilon\bar q}\,\varphi^a\ ,
\]
it follows that

\be
[\,\delta^{}_{J^+}\,,\,\deltab^{int}_{\varepsilon\overline {\mathcal Q}}\,]\,\varphi^a~=~0\ .\ee

\item From

\[
[\,\delta^{}_{\bar\varepsilon q}\,,\,\deltab^{}_{\varepsilon\overline {\mathcal Q}}\,]\,\varphi^a~=-\bar\varepsilon_m\varepsilon^m\,\delta^{}_P\,\varphi^a_{}\ ,\qquad [\,\delta^{}_{\varepsilon \bar q}\,,\,\deltab^{}_{\overline\varepsilon {\mathcal Q}}\,]\,\varphi^a~=~\bar\varepsilon_m\varepsilon^m\,\delta^{}_P\,\varphi^a_{}
\ ,\]
we deduce that 

\be 
[\,\delta^{}_{\bar\varepsilon q}\,,\,\deltab^{int}_{\varepsilon\overline {\mathcal Q}}\,]\,\varphi^a~=~[\,\delta^{}_{\bar\varepsilon q}\,,\,\deltab^{int}_{\overline\varepsilon {\mathcal Q}}\,]\,\varphi^a~=~0\ .
\ee

\item Proper transformation under $J^{+-}$

\be
[\,\delta^{}_{J^{+-}}\,,\,\deltab^{int}_{\varepsilon\overline {\mathcal Q}}\,]\,\varphi^a~=-\frac{i}{2}\,\deltab^{int}_{\varepsilon\overline {\mathcal Q}}\,\varphi^a\ ,\qquad [\,\delta^{}_{J^{+-}}\,,\,\deltab^{int}_{\overline\varepsilon {\mathcal Q}}\,]\,\varphi^a~=-\frac{i}{2}\,\deltab^{int}_{\overline\varepsilon {\mathcal Q}}\,\varphi^a
\ee

\item Dimension analysis requires
\be
[\,\delta^{}_{D}\,,\,\deltab^{int}_{\varepsilon\overline {\mathcal Q}}\,]\,\varphi^a~=~\frac{i}{2}\,\deltab^{int}_{\varepsilon\overline {\mathcal Q}}\,\varphi^a\ ,\qquad [\,\delta^{}_{D}\,,\,\deltab^{int}_{\overline\varepsilon {\mathcal Q}}\,]\,\varphi^a~=~\frac{i}{2}\,\deltab^{int}_{\overline\varepsilon {\mathcal Q}}\,\varphi^a\ .
\ee

\item They have opposite $U(1)$ $R$-charge,

\be
[\,\delta^{}_{J}\,,\,\deltab^{int}_{\varepsilon\overline {\mathcal Q}}\,]\,\varphi^a~=~\frac{1}{2}\,\deltab^{int}_{\varepsilon\overline {\mathcal Q}}\,\varphi^a\ ,\qquad [\,\delta^{}_{J}\,,\,\deltab^{int}_{\overline\varepsilon {\mathcal Q}}\,]\,\varphi^a~=-\frac{1}{2}\,\deltab^{int}_{\overline\varepsilon {\mathcal Q}}\,\varphi^a\ .
\ee

\item The eight interacting supersymmetries must also transform as an $SO(8)$ vector: with $\bar\varepsilon'^{}_m=\omega^{}_{mn}\varepsilon^n_{}$,

\be
[\,\delta^{}_{{coset}}\,,\,\deltab^{}_{\varepsilon\overline {\mathcal Q}}\,]\,\varphi^a~=~\deltab^{}_{ \bar\varepsilon'{\mathcal Q}}\,\varphi^a\ , 
\ee

\be\label{cosetbar}
[\,\delta^{}_{\overline{coset}}\,,\,\deltab^{}_{\varepsilon\overline {\mathcal Q}}\,]\,\varphi^a~=~0\ .
\ee
Similarly, with $\varepsilon'^m=\omega^{mn}\overline\varepsilon_n$,

\be
[\,\delta^{}_{\overline{coset}}\,,\,\deltab^{}_{\overline\varepsilon {\mathcal Q}}\,]\,\varphi^a~=~\deltab^{}_{ \varepsilon'\overline{\mathcal Q}}\,\varphi^a\ , 
\ee

\be\label{cosetbar2}
[\,\delta^{}_{{coset}}\,,\,\deltab^{}_{\overline\varepsilon {\mathcal Q}}\,]\,\varphi^a~=~0\ .
\ee

\end{enumerate}
These ten requirements limit the possible forms of the dynamical supersymmetries.
%%%%%%%%%%%%%%%%%%

\section{Solving the Kinematical Restrictions}
In order to satisfy the first two requirements, we must construct chiral cubic polynomials in the superfields, which requires a bit of algebraic technology.
\subsection{Chiral Engineering}
Introduce the coherent state operators

\be E_{\eta}^{}~=~e^{\,\eta\cdot\widehat{\overline d}}_{}\ ,
%\qquad E^{-1}_{\eta}~=~e^{\,-\eta\cdot\widehat{\overline d}}_{}\ ,
\ee
where the hat denotes division by $\partial^+$, and $\eta^m$ are arbitrary Grassmann parameters. Since  

\be 
d^m_{}\,\left(\,E_{\eta}^{}\,\varphi^a\,\right)~=~ i\sqrt{2}\,\eta^m_{}\,\left(\,E_{\eta}^{}\,\varphi^a \,\right)\ ,
\ee
$E_{\eta}\,\varphi^a$ are eigenstates of the chiral derivatives. It follows that the quadratic combination

\be
Z^{bc(\eta)}_{}~=~
%\frac{1}{\partial^{+A}}\left[\,
(E^{}_{\eta}\partial^{+B}_{}\,\varphi^b_{})\,(E^{}_{-\eta} \partial^{+C}\,\varphi^c)
%\right]
\ ,
%&\equiv&\frac{1}{\partial^{+A}}\left[\,E^{}_{\eta}\partial^{+B}_{}\,\varphi^b_{}\,\cdot\,E^{-1}_{\eta} \partial^{+C}\,\varphi^c\right]
\ee
is manifestly chiral,

\be
d^m\,Z^{bc(\eta)}~=~0\ .
\ee
Chiral cubic polynomials in the superfields are then constructed in {\em nested form},

\be
{\cal C}^{bcd\,(\eta,\zeta)}_{}~=~
%\frac{1}{\partial^{+A}}\left[\,
(E^{}_{\eta}\partial^{+B}_{}\,\varphi^b_{}\,)
%\cdot
\,E^{}_{-\eta} \frac{1}{\partial^{+M}}\left(\,(E^{}_{\zeta}\partial^{+C}\,\varphi^c\,)
%\cdot
(\,E^{}_{-\zeta}\partial^{+D}\,\varphi^d\,)\,\right)
%\right]
\ ,
\ee
which is manifestly chiral

\be
d^m_{}\,{\cal C}^{bcd(\eta,\zeta)}_{}~=~0\ ,\ee
and serves as  a generating function where the chiral cubic polynomials in the superfields appear as the coefficients in the series expansion in the independent Grassmann variables $\eta$ and $\zeta$. 

%%%%%%%%%%%%%
\subsection{Dynamical Supersymmetry}
To find it, we introduce the supersymmetry parameters in the nested Ansatz through the combinations 

\be 
E_{\varepsilon}^{}~=~e^{\,\varepsilon\cdot\widehat{\overline q}}_{}\ ,\qquad  E_{\,\bar\varepsilon}^{}~=~e^{\,\bar\varepsilon\cdot{\widehat q}}\ , \ee
which allows us to keep track of requirement (6), without affecting chirality. This leads to the nested ans\" atze of the form   

\bea
\deltab^{}_{\varepsilon\overline {\mathcal Q}}\,\varphi^a&=&\frac{f^{abcd}_{}}{\partial^{+A_\alpha}}\left((E_{\varepsilon}^{}E_{\eta}^{}\partial^{+B_\alpha}_{}\varphi^b_{})E_{-\varepsilon}^{}E_{-\eta}^{} \frac{1}{\partial^{+M_\alpha}}\left((E_{\zeta}^{}\partial^{+C_\alpha}\varphi^c)(E_{-\zeta}^{}\partial^{+D_\alpha}\varphi^d\,)\right)\right)\ ,\nonumber\\
&\equiv&
{\cal K}^{a\,(\varepsilon,\eta,\zeta)}_\alpha\ ,
\eea
keeping only the first order in the supersymmetry parameters $\varepsilon^m$. The $f^{abcd}$ are unknown coefficients, and the exponents $A_\alpha$, $B_\alpha$, $M_\alpha$, $C_\alpha$, $D_\alpha$ have yet to be determined. In this form, many of the ten requirements are manifestly satisfied:

\begin{itemize}

\item Chirality is manifest
since the $\bar q_n$ anticommute with the chiral derivatives.

\item  Requirements (3), (4), (5), and (6) are clearly satisfied. 

\item The proper transformation under $J^{+-}$, (7), restricts the power of the $\partial^+$ derivatives so that   

\be
A_\alpha^{}+M^{}_\alpha-B^{}_\alpha-C^{}_\alpha-D^{}_\alpha+4~=~0\ ,\ee
which also satisfies the dimension requirement (8).

\item The correct $U(1)$ $R$-charge, requirement (9), demands after some computation

\be\label{nineb}
\left(\,\eta^m\frac{\partial}{\partial\eta^m}+\zeta^m\frac{\partial}{\partial\zeta^m}~-~4\,\right)\,{\cal K}^{a\,(\varepsilon,\eta, \zeta)}_\alpha~=~0 \ .
\ee

\item The tenth requirement, that the eight supersymmetries transform as an $SO(8)$ vector, is the hardest to satisfy. 
Computation of the commutator yields

\bea\nonumber
[\,\delta^{}_{{coset}}\,,\,\deltab^{}_{\varepsilon\overline {\mathcal Q}}\,]\,\varphi^a&=&\deltab^{}_{ \bar\varepsilon'{\mathcal Q}}\,\varphi^a \\
&&+
\,\omega^{}_{mn}\left(\,\eta^m_{}\eta^n_{}(\widehat U_1+\widehat U_2)+\zeta^m\zeta^n(\widehat U_3+\widehat U_4)\,\right){\cal K}^{a\,(\epsilon,\eta,\zeta)}_\alpha
%\,2\,\omega^{}_{mn}(\eta_3^n+\eta_4^n)\eta_2^m\,\hat U_2{\cal K}^{a\,(\vec\varepsilon,\vec\eta)}_\alpha+
%\omega^{}_{mn}\sum_{i=1}^{4}\,\eta^n_i\eta_i^m\,\hat U_i\,{\cal K}^{a\,(\vec\varepsilon,\vec\eta)}_\alpha
\ .\nonumber\\
&&
\eea
Here, $\widehat U_i$ means insertion of $\frac{1}{\partial^{+}}$ in the $i^{th}$ position; for instance 

\bea
&&\widehat U_2 {\cal K}^{a\,(0,\eta,\zeta)}_\alpha=\nonumber \\
&&\frac{f^{abcd}_{}}{\partial^{+A_\alpha}}\left((E_{\eta}^{}\partial^{+B_\alpha}_{}\varphi^b_{})E_{-\eta}^{} \frac{1}{\partial^{+(M_\alpha+1)}}\left((E_{\zeta}^{}\partial^{+C_\alpha}\varphi^c)(E_{-\zeta}^{}\partial^{+D_\alpha}\varphi^d\,)\right)\right)\nonumber \ ,
\eea
and so on. Hence the tenth kinematical requirement of $SO(8)$ covariance is achieved as long as

\be
\label{extra}
\left(\,\eta^m_{}\eta^n_{}(\widehat U_1+\widehat U_2)+\zeta^m\zeta^n(\widehat U_3+\widehat U_4)\,\right){\cal K}^{a\,(\epsilon,\eta,\zeta)}_\alpha~=~0\ .\ee
After some algebra, we find two solutions to this equation.

\vskip .2cm
The {\it odd solution} 
%\[\delta^{{\, int}\, even}_{\varepsilon\overline {\cal Q}}\,\varphi^a~=~\sum_{even}\,{\cal K}^{a\,(\varepsilon,\eta,\zeta)}_\alpha\,\Big|_{\eta=\zeta=0}^{}\] 

\be
\delta^{{\rm int}\, odd}_{\varepsilon\overline {\cal Q}}\,\varphi^a~=~\sum_{odd}\,{\cal K}^{a\,(\varepsilon,\eta,\zeta)}_\alpha\,\Big|_{\eta=\zeta=0}^{}\ ,
%(-1)^\alpha\,\frac{\partial}{\partial\eta^{[2-2\alpha]}}\frac{\partial}{\partial\zeta^{[2+2\alpha]}}\Delta^{(\bar\epsilon\eta,\zeta)}_\alpha
\ee 
where the sum stands for

\be
\sum_{odd}~\equiv \sum_{\alpha=\pm\frac{1}{2}}
(-1)^{\alpha+\frac{1}{2}}\,\frac{\partial}{\partial\eta^{[2-2\alpha]}}\frac{\partial}{\partial\zeta^{[2+2\alpha]}}\ ,\ee
with

\be
\frac{\partial}{\partial\eta^{[2-2\alpha]}}\frac{\partial}{\partial\zeta^{[2+2\alpha]}}~\equiv~
\frac{{\epsilon^{i_1\cdots i_{2-2\alpha}\cdots i_4}_{}}}{\scriptstyle(2+2\alpha)!(2-2\alpha)!}\frac{\partial}{\partial\eta^{i_1\cdots i_{2-2\alpha}}}\frac{\partial}{\partial\zeta^{i_{3-2\alpha}\cdots i_4}}\ .
\ee

\vskip .2cm
The second is the {\it even solution} 

\be
\delta^{{\rm int}\, even}_{\varepsilon\overline {\cal Q}}\,\varphi^a~=~\,\sum_{even}\,{\cal K}^{a\,(\varepsilon,\eta,\zeta)}_\alpha\,\Big|_{\eta=\zeta=0}^{}
%(-1)^\alpha\,\frac{\partial}{\partial\eta^{[2-2\alpha]}}\frac{\partial}{\partial\zeta^{[2+2\alpha]}}\Delta^{(\bar\epsilon\eta,\zeta)}_\alpha
\ ,\ee 
with 

\be
\sum_{even}~\equiv~\sum_{\alpha=0, \pm 1}
(-1)^\alpha\,\frac{\partial}{\partial\eta^{[2-2\alpha]}}\frac{\partial}{\partial\zeta^{[2+2\alpha]}}\ .\ee
%For both

%\be\vec\eta~=~(\eta,-\eta,\zeta,-\zeta)\ ,\ee
\vskip .2cm
In both cases, the powers of the $\partial^+$ derivatives are related by 

\be
A^{}_{\alpha-1}=A^{}_\alpha+1\ ,\quad B^{}_{\alpha-1}=B^{}_\alpha+1\ ,\quad M^{}_{\alpha-1}=M^{}_\alpha-2\ ,\ee
as well as

\be
C^{}_{\alpha-1}=C^{}_\alpha-1\ ,\quad D^{}_{\alpha-1}=D^{}_\alpha-1
\ .\ee
Both even and odd solutions are seen to satisfy Eq.(\ref{cosetbar}). Their forms suggest that $SO(8)$ triality is at work, with $\alpha$ denoting the $U(1)$ charges in its vector and spinor representations. 

Both solutions are conveniently written in the form 

\be
\delta^{{\, int}\, odd(even)}_{\varepsilon\overline {\cal Q}}\,\varphi^a~\equiv~\bigl[\,A^{}_\alpha,B^{}_\alpha,M^{}_\alpha,C^{}_\alpha,D^{}_\alpha\,\bigr]^{}_{\,odd(even)}
\ ,\ee
with $\alpha=-1/2(-1)$ in the odd(even) case. 

\end{itemize}

It can be checked that these two solutions satisfy the correct commutations with the kinematical conformal supersymmetries

\be
[\,\delta^{}_{\bar\alpha s}\,,\,\deltab^{}_{\varepsilon\overline {\mathcal Q}}\,]\,\varphi^a~=~\left(\,i\delta^{}_{D}\,\varphi^a_{}-i\delta^{}_{J^{+-}}\,\varphi^a_{}+\frac{1}{2}\,\delta^{}_{J}\,\varphi^a_{}\,\right)
+\frac{1}{\sqrt{2}}\,\delta^{}_{SO(6)}\,\varphi^a\ ,
\ee
as well as

\be
[\,\delta^{}_{\alpha\bar s}\,,\,\deltab^{}_{\varepsilon\overline{\mathcal Q}}\,]\,\varphi^a~=~\delta^{}_{\overline{coset}}\,\varphi^a\ .
\ee
 %\[[\,\delta^{}_{SO(6)}\,,\,\deltab^{}_{\overline {\mathcal Q}}\,]\,\varphi^a~=~\deltab^{}_{\bar {\mathcal Q}}\,\varphi^a\]
Finally, we note that the conjugate supersymmetries are obtained by simply changing $E_{\varepsilon}$ into $E_{\bar\varepsilon}$.

%%%%%%%%%%%%%%%%%%%%%%%%%%%%%
\subsection{ Hamiltonian and Boost} 
In the previous section, the form of the dynamical supersymmetry transformations have been narrowed down to two solutions with yet undetermined powers of the light-cone derivatives. In the $d=4$ SuperYang-Mills case, their values were determined from the vanishing of the commutator between the light-cone boost and Hamiltonian. We expect the same to hold in the $d=3$ theory.

The light-cone Hamiltonian is computed from  

\be
[\,\deltab^{free}_{\bar\varepsilon{\cal Q}}+\deltab^{\, int}_{\bar\varepsilon{\cal Q}}\,,\,\deltab^{\, free}_{\varepsilon\overline{\cal Q}}+\deltab^{\, int}_{\varepsilon\overline{\cal Q}}\,]\varphi^a_{}
~=~\sqrt{2}\,\bar\varepsilon^{}_m\varepsilon_{}^m\,\deltab^{\,\rm }_{{\cal P}^-}\,\varphi^a\ .
\ee
The  commutator yields terms linear and quadratic in $f^{abcd}$. The first order stems from

\be
[\,\deltab^{free}_{\bar\varepsilon{\cal Q}}\,,\,\deltab^{\, int}_{\varepsilon\overline{\cal Q}}\,]\varphi^a_{}+
[\,\deltab^{\, int}_{\bar\varepsilon{\cal Q}}\,,\,\deltab^{\, free}_{\varepsilon\overline{\cal Q}}\,]\varphi^a_{}
\ee
The results of the computation are expressed in terms of
  
\bea
K^{a\,[r,1]}_\alpha&\equiv&(E^{}_r\,U^{}_1)(E^{}_{-r}\,U_2)\,{\cal K}^{a\,(0,\eta,\zeta)}_\alpha\ ,\nonumber\\ 
&=&\frac{f^{abcd}_{}}{\partial^{+A_\alpha}}\left((E_{r}^{}E_{\eta}^{}\partial^{+B_\alpha}_{}\varphi^b_{})E_{-r}^{}E_{-\eta}^{} 
\frac{1}{\partial^{+M_\alpha}}\left((E_{\zeta}^{}\partial^{+C_\alpha}\varphi^c)(E_{-\zeta}^{}\partial^{+D_\alpha}\varphi^d\,)\right)\right)\nonumber \ ,
\eea
and

\bea
K^{a\,[1,r]}_\alpha&\equiv&(E^{}_r\,U^{}_3)(E^{}_{-r}\,U_4)\,{\cal K}^{a\,(0,\eta,\zeta)}_\alpha\ ,\nonumber\\
&=&\frac{f^{abcd}_{}}{\partial^{+A_\alpha}}\left((E_{\eta}^{}\partial^{+B_\alpha}_{}\varphi^b_{})E_{-\eta}^{} 
\frac{1}{\partial^{+M_\alpha}}\left((E_{r}^{}E_{\zeta}^{}\partial^{+C_\alpha}\varphi^c)(E_{-r}^{}E_{-\zeta}^{}\partial^{+D_\alpha}\varphi^d\,)\right)\right)\nonumber \ ,
\eea
where the transverse derivative is introduced through

\be
E_{r}^{}~=~e^{\,r\,\widehat\partial}_{} \ ,
\ee
and $r$ is a dimensionless parameter. 

The computation of these commutators yields, for the odd case, 

\be
\deltab^{{\,int}\,odd}_{{\cal P}^-}\,\varphi^a~=~
\frac{\partial}{\partial r}\,
\left(\sum_{even}K^{a\,[1,r]}_\alpha+\sum_{odd}\,K^{a\,[r,1]}_{\alpha+\frac{1}{2}}\right)_{r=0}\ ,
\ee
for the linear part in $f^{abcd}$. 

A similar expression is found in the even case,
 
\be
\deltab^{{\,int}\,even}_{{\cal P}^-}\,\varphi^a~=~
\frac{\partial}{\partial r}\,
\left(\sum_{odd}K^{a\,[1,r]}_\alpha-\sum_{even}\,K^{a\,[r,1]}_{\alpha+\frac{1}{2}}\right)_{r=0}\ .
\ee
The boost transformation is computed from the commutator 

\be
\deltab^{\,\rm }_{{\cal J}^-}\,\varphi^a~=-\frac{i}{2}\,[\,\delta^{}_{K}\,,\,\deltab^{}_{{\cal P}^-}\,]\,\varphi^a_{}\ .
\ee
for both odd and even cases. Its expression is not particularly enlightening, and will be published elsewhere\cite{US}{}. 

\subsection{Dynamical Constraints}
The next step is to require 

\be
[\,\delta^{}_{{\cal P}^-}\,,\,\deltab^{}_{{\cal J}^-}\,]\,\varphi^a_{}
~=~0\ .
\ee
This condition, as in the Yang-Mills case, is expected to fix the unknown exponents, and  the interactions. After a lengthy calculation, keeping only the terms linear in $f^{abcd}$, the result can be written in the form

\be 
[\,\delta^{^{\,odd}}_{{\cal P}^-}\,,\,\deltab^{^{\,odd}}_{{\cal J}^-}\,]\,\varphi^a_{}
~=~{\cal S}\,\frac{\partial^2}{\partial r\partial r'}\,\left(F\,{\cal O}^{{\,odd}}_1+ G\,{\cal O}^{\,odd}_2\right)_{r=r'=0}\ , 
\ee
where $\cal S$ is a shift operator

\be
{\cal S}:~~~ A_\alpha\rightarrow A_\alpha+1\ ,~~B_\alpha\rightarrow B_\alpha+1\ ,~~M_\alpha\rightarrow M_\alpha-1\ .
\ee 
In addition, 

\bea
F&\equiv& (B^{}_{-\frac{1}{2}}-3)\widehat U^{}_1+(M_{-\frac{1}{2}}-C_{-\frac{1}{2}}-D_{-\frac{1}{2}}+3)\widehat U^{}_2\ ,\\
G&\equiv& 2(C^{}_{-\frac{1}{2}}-\frac{3}{2})\widehat U^{}_3-2(D_{-\frac{1}{2}}-\frac{3}{2})\widehat U^{}_4\ ,
\eea
and

\bea
{\cal O}^{\,odd}_1&=& \sum_{odd}( K^{[rr',1]}_{\alpha}-K^{[rr',1]}_{\alpha+1}) +2\sum_{even} K^{[r,r']}_{\alpha+\frac{1}{2}}\ ,
\\
{\cal O}^{\,odd}_2&=& \sum_{odd} K^{[r,r']}_{\alpha+1}-\frac{1}{2}\sum_{even}( K^{[r,r']}_{\alpha+\frac{1}{2}}-K^{[1,rr']}_{\alpha+\frac{3}{2}})\ .
\eea
A similar equation obtains in the even case. 

We can envisage two types of solutions to the vanishing of this commutator. 

\begin{itemize}

\item The ``trivial" solution is when $F=G=0$, which determines the values of all the exponents, and leads to   

\be
\delta^{{\,{\rm int}\,odd}}_{\varepsilon\overline {\cal Q}}\,\varphi^a~=~\bigl[\,2,3,0,\frac{3}{2},\frac{3}{2}\,\bigr]^{}_{\,odd}\ ,
\ee
which exists only if $f^{abcd}=-f^{abdc}$. 

The even solution, given by,  
\be
\delta^{\, {\rm int}\,even }_{\varepsilon\overline {\cal Q}}\,\varphi^a~=~\bigl[\,\frac{5}{2},\frac{7}{2},-1,1,1\,\bigr]^{}_{\,even}\ ,
\ee
requires $f^{abcd}=+f^{abdc}$. 

In both cases, there are no further symmetry requirements on $f^{abcd}$. Both solutions require fractional powers of $\partial^+$, which have to be further interpreted.

We have not checked the validity of this solution any further: there remains to check the vanishing of the  commutators

\be
 [\,\delta^{}_{K}\,,\,\deltab^{}_{{\cal K}^-}\,]\,\varphi^a_{}\ ,\qquad [\,\deltab^{}_{{\cal P}^-}\,,\,\deltab^{}_{{\cal K}^-}\,]\,\varphi^a_{}\ ,\qquad  [\,\delta^{}_{{\cal J}^-}\,,\,\deltab^{}_{{\cal K}^-}\,]\,\varphi^a_{}\ .
 \ee
In the Yang-Mills case, these did not put any further restrictions on the solution. The vanishing of the second one may be explained by the Jacobi identity:

\be
[\,\deltab^{}_{\cal {P}^-}\,,\, [\,\delta^{}_{K}\,,\,\deltab^{}_{\mathcal{J}^-}\,]\,]+
[\,\delta^{}_{K}\,,\, [\,\deltab^{}_{\mathcal{J}^-}\,,\,\deltab^{}_{\cal {P}^-}\,]\,]+
[\,\deltab^{}_{\mathcal{J}^-}\,,\, [\,\deltab^{}_{\cal {P}^-}\,,\,\delta^{}_{K}\,]\,]~=~0\ .
\ee
Since

\be
[\,\delta^{}_{K}\,,\,\deltab^{}_{\mathcal{J}^-}\,]\,\varphi^a_{} ~=-i\,\deltab^{}_{\mathcal{K}^-}\,\varphi^a_{}\ ,
\ee
it follows that 

\be
[\,\deltab^{}_{\cal {P}^-}\,,\,\deltab^{}_{\mathcal{J}^-}\,]\,\varphi^a_{} ~=~0\ .\ee
The algebraic validity of the fractional power solution hinges on the first commutator

\be
 [\,\delta^{}_{K}\,,\,\deltab^{}_{\mathcal{K}^-}\,]\,\varphi^a_{} ~=~0\ ,
 \ee 
which we have not yet checked. Through the Jacobi identity, it would ensure that 

\be
 [\,\deltab^{}_{\mathcal{J}^-}\,,\,\deltab^{}_{\mathcal{K}^-}\,]\,\varphi^a_{} ~=~0\ .
\ee

\item The less trivial solution(s) relies on the symmetries of $f^{abcd}$ under the interchange of {\em three} of its indices. It appears to lead uniquely to the BLG solution; since at the time of Shifmania, we had not obtained it, its details will appear elsewhere\cite{US}{}. 

\end{itemize}

Much remains to be done. For one, we have not derived the quadratic term in $f^{abcd}$ in the Hamiltonian. In the Yang-Mills case, this led to the Jacobi identity of the $f^{abc}$ and identified them as structure functions.    

\section*{Acknowledgments}
It has been an honor to be invited to speak at Misha's celebration, as well as a pleasure to attend Shifmania, where Misha's superb skills and influence 
were so deservedly lauded, and the wonderful Russian hospitality was much in evidence.

This research is partially supported by the Department of Energy Grant No. DE-FG02-97ER41029.

%%%%%%%%%%%%

\end{document}